\documentclass[aps,prb,twocolumn,showpacs,amsmath,amssymb]{revtex4}
\usepackage{epsfig}
\usepackage{bm}
\def\be{\begin{equation}}
\def\ee{\end{equation}}

\begin{document}

\title{Electrical current noise of a beam splitter as a test of
spin-entanglement} \author{P. Samuelsson, E.V. Sukhorukov and
  M. B\"uttiker} 
\affiliation{D\'epartement de
Physique Th\'eorique, Universit\'e de Gen\`eve, CH-1211 Gen\`eve 4,
Switzerland} \date{\today}

\begin{abstract}
We investigate the spin entanglement in the superconductor-quantum dot
system proposed by Recher, Sukhorukov and Loss, coupling it to an
electronic beam-splitter. The superconductor-quantum dot entangler and
the beam-splitter are treated within a unified framework and the
entanglement is detected via current correlations. The state emitted
by the entangler is found to be a linear superposition of non-local
spin-singlets at different energies, a spin-entangled two-particle
wavepacket. Colliding the two electrons in the beam-splitter, the
singlet spin-state gives rise to a bunching behavior, detectable via
the current correlators. The amount of bunching depends on the
relative positions of the single particle levels in the quantum dots
and the scattering amplitudes of the beam-splitter. It is found that
the bunching-dependent part of the current correlations is of the same
magnitude as the part insensitive to bunching, making an experimental
detection of the entanglement feasible. The spin entanglement is
insensitive to orbital dephasing but suppressed by spin dephasing. A
lower bound for the concurrence, conveniently expressed in terms of
the Fano factors, is derived. A detailed comparison between the
current correlations of the non-local spin-singlet state and other
states, possibly emitted by the entangler, is performed. This provides
conditions for an unambiguous identification of the non-local singlet
spin entanglement.
\end{abstract}
\maketitle 
\section{Introduction}

Ever since the concept of entanglement was introduced, \cite{Schrod}
it has been at the heart of conceptual discussions in quantum
mechanics.  \cite{EPR,Bohr,Bohm} The discussions have mainly concerned
the non-local properties of entanglement. Two entangled, spatially
separated particles, an Einstein-Podolsky-Rosen \cite{EPR} (EPR) pair,
are correlated in a way which can not be described by a local,
realistic theory, i.e. the correlations give rise to a violation of a
Bell Inequality.\cite{Bell} In optics, the non-local properties of
entangled pairs of photons have been intensively investigated over the
last decades.\cite{Zeil,Gisin98,Weihs} Recently, the interest has
turned to possible applications making use of the properties of
entangled particles. Entanglement plays an important role in many
quantum computation and information schemes,\cite{Nielsen} with
quantum cryptography\cite{Gisin02} and quantum
teleportation\cite{Bennet,exptel} as prominent examples.

Compared to optics, the investigation of entanglement in solid state
systems is only in it's infancy. However, a controlled creation,
manipulation and detection of entanglement is a pre-requisite for a
large-scale implementation of quantum computation and information
schemes, making it of large interest to pursue the investigation of
entanglement in solid state systems. Considerable experimental
\cite{Nakamura} and theoretical \cite{Schon} progress has already been
made in the understanding of entangled qubits implemented with
Josephson junctions.

For the entanglement of individual electrons, recently several
important steps towards an experimental realization in mesoscopic
conductors were taken. A scheme for entanglement of orbital degrees of
freedom was proposed in Ref. [\onlinecite{Sam1}], allowing for control
of the entanglement with experimentally accessible electronic
beam-splitters.  \cite{Oliver,Schonenberger} Moreover, several
proposals \cite{Maitre,BI,Sam1} for detecting entanglement via a
violation of a Bell Inequality, expressed in terms of zero-frequency
noise correlators, \cite{Blanter} have been put forth. Very recently,
following a proposal by Beenakker {\it et al} [\onlinecite{Been1}],
several works have discussed the possibility of electron-hole and
post-selected electron-electron entanglement. \cite{noint,samHBT} In
particular, entanglement in the electrical analog of the optical
Hanbury Brown Twiss effect \cite{HBT} was investigated in a mesoscopic
conductor in the quantum Hall regime, transporting electrons along
single edge-states and using quantum point contacts as
beam-splitters.\cite{samHBT} Moreover, a scheme for energy-time
entanglement\cite{Franson} has been proposed.\cite{Gisin03} The
consequences of dephasing for orbital entanglement have been
investigated \cite{Sam1,dephase,Samdephase} as well.

Earlier proposals for electronic entanglement have been based on
creating and manipulating spin entanglement, in normal
\cite{spinentn,Burk00,Egues} as well as in
normal-superconducting\cite{Recher,spinentns,Recher2} systems. Spins
in semiconductors have been shown\cite{Awshalom} to have dephasing
times approaching microseconds, making spins promising candidates for
carriers of quantum information. However, a direct detection of spin
entanglement in mesoscopic conductors is difficult. The natural
quantity to measure is the electrical charge current. To investigate
spin current, one thus in principle has to convert the spin current to
charge current via e.g. spin-filters. Although efficient spin-filters
\cite{spinfilt} have very recently been realized
experimentally,\cite{Kov} there are considerable remaining
experimental complications in manipulating and detecting individual
spins on a mesoscopic scale. In particular, to detect the entanglement
by a violation of a Bell Inequality, one needs \cite{BI} two spin
filters with independent and locally controllable directions to mimic
the polarizers in optical schemes. \cite{Zeil,Gisin98,Weihs}

An alternative idea to detect spin entanglement was proposed by
Burkard, Loss and Sukhorukov [\onlinecite{Burk00}] and also discussed
qualitatively by Oliver, Yamaguchi and Yamamoto
[\onlinecite{Maitre}]. They proposed to use the relation between the
spin and orbital part of the wavefunction, imposed by the antisymmetry
of the total wavefunction under exchange of two particles. A state
with an antisymmetric, singlet spin wavefunction has a symmetric
orbital wavefunction and vice versa for the spin triplet. When
colliding the electrons in a beam-splitter, spin singlets and triplets
show a bunching and anti-bunching behavior respectively. These
different bunching behaviors were found to be detectable via the
electrical current correlations, i.e. the properties of the orbital
wavefunction were used to deduce information about the spin
state. This approach was later extended to all moments of the
current.\cite{Taddei} Moreover, it was recently further elaborated in
Ref. [\onlinecite{Burk03}], taking spin dephasing and non-ideal
beam-splitters into account.

In comparison to detecting spin entanglement via a violation of a Bell
Inequality, the approach of Ref. [\onlinecite{Burk00}] however has a
fundamental limitation. The antisymmetric spin singlet is an entangled
state, while symmetric, triplet spin states are not neccesarily
entangled. Considering e.g. the standard singlet-triplet basis, only
one of the three triplets
$|\uparrow\downarrow\rangle+|\downarrow\uparrow\rangle,
~|\uparrow\uparrow\rangle$ and $|\downarrow\downarrow\rangle$ is spin
entangled. However, all spin-triplet states, having the same
symmetrical orbital wavefunction, give rise to the same anti-bunching
behavior in the current correlators.\cite{Burk00} As a consequence, in
contrast to a Bell Inequality test, the approach of
Ref. [\onlinecite{Burk00}] can not be employed to distinguish between
entangled and non-entangled triplet states. To be able to distinguish
between different triplet state, one would need to consider more
involved schemes, implementing in addition e.g. single spin
rotations. \cite{Egues}

Despite this fundamental limitation, the approach of
Ref. [\onlinecite{Burk00}] is due to its comparable simplicity
still of interest for entanglers emitting non-local
spin-singlets. However, the investigations in
Ref. [\onlinecite{Burk00}] were carried out assuming a
discrete spectrum of the electrons and a mono-energetic entangled
state incident on the beam-splitter. While giving a qualitatively
correct picture of the physics, it does not quantitatively describe
the situation in a conductor connected to electronic reservoirs, where
the spectrum is continuous and the entangled electrons generally have
a wave-packet nature, i.e. the wavefunction is a linear superposition
of entangled electrons at different energies. \cite{Sam1} Moreover,
the wavefunction in Ref. [\onlinecite{Burk00}] was not derived
considering a specific entangler, it was instead taken to be an
incoming plane wave with unity amplitude. This makes the calculated
current correlations inapplicable to most of the entanglers considered
theoretically, \cite{spinentn,Recher,spinentns} which operate in the
tunneling regime and emit entangled states with a low amplitude.
\begin{figure}[h]
\centerline{\psfig{figure=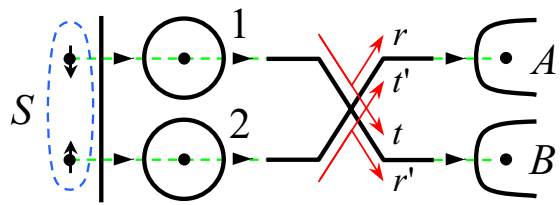,width=0.9\linewidth}}
\caption{Schematic picture of the system. A superconductor (S) is
connected, via tunnel barriers to two quantum dots (1 and
2) in the Coulomb blockade regime. The dots are further coupled, via a
second pair of tunnel barriers, to normal leads which cross in a
forward scattering single-mode beam-splitter. The beam-splitter is
characterized by scattering amplitudes $r,t,r'$ and $t'$. On the other
side of the beam-splitter, the normal leads are connected to normal
electron reservoirs A and B.}
\label{fig1}
\end{figure}

In this paper, we revisit the approach of detection of spin-singlet
entanglement presented in Ref. [\onlinecite{Burk00}]. The
abovementioned shortcomings are bypassed by treating the entangler and
the beam-splitter within a unified theoretical framework. As a source
of non-local spin-singlets, the superconductor-quantum dot entangler
(see Fig. \ref{fig1}) investigated in detail by Recher, Sukhorukov and
Loss in Ref. [\onlinecite{Recher}], is considered. Using a formal
scattering approach, the wavefunction of the electrons emitted from
the entangler is calculated. It is found to be a linear superposition
of pairs of spin-entangled electrons at different energies, a two
electron wavepacket, similar to what was found for the superconducting
orbital entangler in Ref. [\onlinecite{Sam1}]. The amplitude at each
energy depends on the position of the single particle levels in the
dots. Both the process where the electrons tunnel through different
dots, creating the desired non-local EPR-pair, as well as the unwanted
process when both electrons tunnel through the same dot, are
investigated. In both cases the spin wavefunction is a singlet,
preserving the spin-state of the Cooper pair tunneling out of the
superconductor, however the orbital states are different.

The electrons emitted by the entangler are then collided in a
beam-splitter and detected in two electronic reservoirs. Due to the
singlet spin state, electrons tunneling through different dots show a
bunching behavior when colliding in the beam-splitter. Both the auto
and cross correlations between currents flowing into the normal
reservoirs (but not the average current) depend on the degree of
bunching. We find that the bunching is proportional to the
wavefunction overlap of the two colliding electrons. This overlap
depends strongly on the position of the single-particle levels in the
dot, being maximal for both levels aligned with the chemical potential
of the superconductors. The part of the current correlators sensitive
to bunching is of the same magnitude as the part insensitive to
bunching, making an experimental detection of the spin-singlet
entanglement feasible.

The current correlators are independent of scattering phases and thus
insensitive to orbital dephasing. However spin dephasing generally
leads to a mixed spin state with a finite fraction of triplets. Since
the spin triplets have a tendency to anti-bunch, the spin dephasing
results in a reduction of the overall bunching behavior and
eventually, for strong spin-dephasing, to a cross-over to an
anti-bunching behavior. A simple expression for the concurrence,
quantifying the entanglement in the presence of spin dephasing, is
derived in terms of the Fano factors.

For electrons tunneling through the same dot, the wavefunction is a
linear superposition of states for the pair tunneling through dots 1
and 2.  Both the cross- and auto correlators contain a two-particle
interference term, sensitive to the position of the single-particle
levels in the dots, however in a different way than the bunching
dependent term for tunneling through different dots.  In particular,
the correlators depend on the scattering phases, providing a way to
distinguish between the two tunneling processes by modulating e.g. the
the Aharonov-Bohm phase.\cite{Recher} Moreover, the phase dependence
makes the correlators sensitive to orbital dephasing, while the spin
part of the wavefunction is insensitive to dephasing.

\section{The superconductor-quantum dot entangler}
A schematic picture of the system is shown in Fig. \ref{fig1}. A
superconducting (S) electrode is connected to quantum dots (1 and 2)
via tunnel barriers. The dots are further contacted, via normal leads
to a controllable single-mode electronic beam-splitter \cite{Oliver}
characterized by the forward scattering amplitudes $r,t,r'$ and
$t'$. The arms going out from the beam-splitter are connected to
normal electron reservoirs A and B.
\begin{figure}[h]
\centerline{\psfig{figure=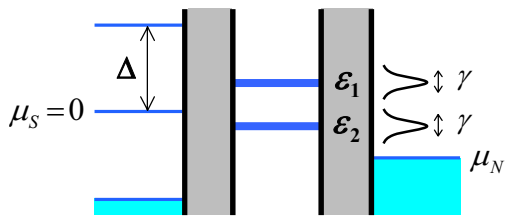,width=0.8\linewidth}}
\caption{Energy diagram of the entangler-beam-splitter system in
  Fig. \ref{fig1}. A bias $eV$ is applied between the superconducting
  reservoir, with chemical potential $\mu_S=0$, and the normal
  reservoirs $A$ and $B$, with the same chemical potential
  $\mu_N=-eV$. There is only one spin-degenerate level of each dot,
  with energy $\varepsilon_1$ and $\varepsilon_2$ respectively, in the
  energy range $-eV$ to $eV$. The level width $\gamma$ is determined by the
  coupling to the normal reservoirs. The bias $eV$ is taken to be much
  smaller than the superconducting gap, $eV \ll \Delta$, but so large
  that the broadened levels are well within the bias window,
  $eV-|\varepsilon_j|\gg \gamma$, $j=1,2$.}
\label{fig1b}
\end{figure}

We first concentrate on a description of the entangler, the
superconductor-quantum dot part of the structure in Fig. \ref{fig1},
investigated in great detail in Ref. [\onlinecite{Recher}]. The
entangler was also recently examined within a density matrix
approach. \cite{Sauret} The role of the beam-splitter is discussed
further below, after a discussion of the quantum state emitted by the
entangler. To simplify our presentation we carry over the notation
from Ref. [\onlinecite{Recher}] when nothing else is stated.

An energy diagram of the superconductor-quantum dot-normal lead part
of the structure is shown in Fig. \ref{fig1b}. A negative bias $-eV$
is applied to the normal reservoirs while the superconductor is
grounded. The chemical potential of the superconductor is taken as a
reference energy, $\mu_S=0$, giving the chemical potential of both
normal reservoirs $\mu_{NA}=\mu_{NB}\equiv \mu_N=-eV$. Each dot 1 and
2 contain a single, spin-degenerate level in the energy range $-eV$ to
$eV$, with energy $\varepsilon_1$ and $\varepsilon_2$
respectively. The level spacing in the dots is assumed to be much
larger than the applied bias, so no other levels of the dots
participate in the transport. The temperature is much lower than the
applied bias (but much larger than the Kondo temperature).

The tunnel barriers between the dots and the superconductor are much
stronger than the tunnel barriers between the dots and the normal
leads. As a consequence, the broadening $\gamma$ of the levels in the
dots (taken the same for both dots) results entirely from the coupling
to the normal leads. The voltage is applied such that the entire
broadened resonances are well within the bias window,
i.e. $eV-|\varepsilon_j|\gg \gamma$ with $j=1,2$. The quantum dots are
in the Coulomb blockade regime, i.e. it costs a charging energy $U$ to
put two electrons on the same dot. The ground state
contains an even number of electrons in the lower lying levels,
i.e. anti-ferromagnetic filling of the dots.
\begin{figure}[h]
\centerline{\psfig{figure=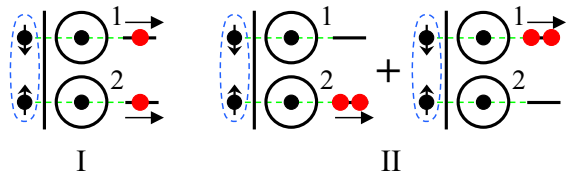,width=0.9\linewidth}}
\caption{Tunneling processes transporting two electrons from the
  superconductor to the normal leads. Process I, where the two electrons tunnel
  through different dots, one through dot 1 and one through dot 2,
  creates the wanted EPR-pair. Process II, where
  the two electrons tunnel through the same dot, either both through
  dot 1 or both through dot 2, is unwanted.}
\label{fig1c}
\end{figure}

The transport takes place as Cooper pairs tunnel from the
superconductor, through the dots and out into the normal leads. Due to
the dominating tunnel barrier at the dot-superconductor interface, one
pair that tunneled onto the dots leaves the dots well before the next
pair tunnels. There are two distinct possibilities for the Cooper pair
to tunnel from the superconductor to the normal leads, shown in
Fig. \ref{fig1c}: 
\begin{itemize} 
\item I, the pair
splits and one electron tunnels through each dot, 1 and 2.
\item II, both
electrons tunnel through the same dot, 1 or 2.
\end{itemize}
It was shown in Ref. [\onlinecite{Recher}] that under the conditions
stated above, all other tunneling processes could be neglected. The
process I creates the wanted EPR-pair, a spin singlet state with the
two electrons spatially separated. However, in an experiment one can
not a priori exclude the second, unwanted process, II. One thus has to
investigate process II as well, to provide criteria for an unambiguous
experimental identification of emission of EPR-pairs.

The first process, I, with the two electrons tunneling through
different dots, is suppressed below the single particle tunneling
probability squared, since the two electrons have to leave the
superconductor from two spatially separated points, i.e. effectively
breaking up the Cooper pair. The tunneling amplitude for a ballistic,
three dimensional superconductor\cite{cooperbreak,Recher} is $A_0
\propto \mbox{exp}(-d/\xi)/(k_{FS}d)$ where $d$ is the distance
between the superconductor-dot connection points, $k_{FS}$ the Fermi
wave number in the superconductor and $\xi$ the superconducting
coherence length. This amplitude is in general larger for lower
dimensional \cite{Recher2} and disordered \cite{Fein,Pist}
superconductors. An investigation of the dependence of $A_0$ on the
geometry of the contacts to the superconductor was performed in
Refs. [\onlinecite{sols,Pist}]. We point out that ways to avoid the
suppression due to pair breaking by means of additional dots have been
discussed in a similar context in
Ref. [\onlinecite{Sanchez}]. However, since more dots complicate the
calculation as well as the experimental realization, we consider the
simpler geometry in Fig. \ref{fig1}.

The second process, II, with both electrons tunneling through the same
dot, is suppressed by the Coulomb blockade in the dots, as $1/U$. In
addition, there is a process which avoids double occupancy of the dots
but instead requires a pair breaking, leading to suppression of the
order $1/\Delta$. Together, this gives an amplitude $B_0\propto
(1/U+1/\Delta \pi)$. The exact expression for the constants $A_0$ and
$B_0$ in terms of tunnel amplitudes between the dots and the
superconductor and the dots and the leads can be found in
Ref. [\onlinecite{Recher}], for our purposes these expressions are
not necessary.

We point out that possibles candidates for experimental realization of
the proposed system are the extensively investigated\cite{Schapers}
heterostructures with semiconductors contacted to metallic
superconducting electrodes. Electron transport through double dots in
semiconductor systems have been recently been reviewed, \cite{Kouwrew}
with an emphasis on experimental advances.

\section{The wavefunction of the spin-entangled electrons.} 

To calculate the wavefunction of the electrons emitted from the
superconductor-quantum dot entangler, we employ the formal scattering
theory \cite{Merzbacher} with the Lippman-Schwinger equation expressed
in terms of the transfer matrix (T-matrix). The total Hamiltonian of
the system can be written as $H=H_0+H_T$, where $H_0$ is the
Hamiltonian of the superconductor, the quantum dots, and the normal
leads. The perturbation $H_T$ describes tunneling between the
superconductor, dots, and leads.  The exact many-particle state
$|\Psi\rangle$ satisfies the Schr{\"o}dinger equation
$(E-H)|\Psi\rangle=0$. In the absence of a perturbation, $H_T=0$, the
system is in the ground state
$|0\rangle=|0\rangle_S|0\rangle_D|0\rangle_N$, with different chemical
potentials, $\mu_S=0$, and $\mu_N=-eV$. The perturbation $H_T$ causes
the electrons to tunnel from the superconductor, via the quantum dots,
to the normal leads.

We use the local nature of the tunneling perturbation and take the
formal scattering approach to the problem. According to this approach
the state $|\Psi\rangle$ can be obtained by solving the
Lippman-Schwinger equation in Fock-space
\begin{equation}
|\Psi\rangle=|0\rangle + \hat G(0)H_T|\Psi\rangle,
\label{LS}
\end{equation}
where the retarded operator $\hat G(E)=[E-H_0+i0]^{-1}$ gives a state
describing particles going out from the scattering region. Note that
the total energy of the ground state $|0\rangle$ is $E=0$. The formal
solution of Eq.\ (\ref{LS}) can be written as
\begin{equation}
|\Psi\rangle=|0\rangle + \hat G(0)T(0)|0\rangle,
\end{equation}
where 
\begin{equation}
T(E)=H_T+H_T\sum_{n=1}^{\infty}[\hat G(E)H_T]^n
\label{T-matrix}
\end{equation}
is the $T$-matrix. One then inserts a complete set of many body states
$1=\sum_N|N\rangle\langle N|$ with $|N\rangle$ the eigenbasis of the
Hamiltonian $H_0$, i.e. the basis of Fock-states of electrons and
quasiparticles in the leads, dots and superconductor respectively. The
quantum number $N$ collectively denotes the energies, spins, lead and
dot indices etc of the individual particles. The eigenenergy of the
state $|N\rangle$, i.e. the total energy of the individual particles,
is $E_N$. This gives an expression for the state
\begin{equation}
|\Psi\rangle=|0\rangle - \sum_N\frac{1}{E_N-i0}|N\rangle\langle N|T(0)|0\rangle.
\label{LS-solution1}
\end{equation}
In the system under consideration, all relevant matrix
elements\cite{Recher} $\langle N|T(0)|0\rangle$ are analytic in the
upper part of the complex energy plane. As a consequency, in the
integration over energies of the individual particles in $|N\rangle$,
the pole arising from the denominator $E_N-i0$ can be replaced by a
$\delta(E_N)$-function, imposing a total energy $E_N=0$, equal to the
chemical potential energy of the superconductor. This gives the
wavefunction
\begin{equation}
|\Psi\rangle=|0\rangle-2\pi i
\sum_N\delta(E_N)|N\rangle\langle N|T(0)|0\rangle.
\label{LS-solution2}
\end{equation}  
It was shown in Ref.\ [\onlinecite{Recher}], that under the conditions
stated above and to lowest order in coupling between the
superconductor and the dots, the operator $T$ creates from the vacuum
$|0\rangle$ a two-electron spin-entangled state. As pointed out above,
depending on the relation between the amplitudes $A_0$ and $B_0$, the
transport of the two electrons through the same (process II) or
different (process I) dots dominates. Below we consider for simplicity
only the limiting cases, where either I or II is completely
dominating, however our analysis can straightforwardly be extended to
a situation where they are of comparable strength.

\subsection{Electrons tunneling through different dots.}

We first consider process I, when the amplitude for tunneling through
different dots is much larger than the amplitude to tunnel through the
same dot. This creates the desired EPR-pair, a non-local
spin-entangled pair of electrons. The quantities in this limit are
denoted with a I. The wavefunction for two spin-entangled electrons at
energies $E_1$ and $E_2$ is
\begin{equation}
|E_1,E_2\rangle_I=
\frac{1}{\sqrt{2}}[b_{1\uparrow}^{\dagger}(E_1)b_{2\downarrow}^{\dagger}(E_2)
-b_{1\downarrow}^{\dagger}(E_1)b_{2\uparrow}^{\dagger}(E_2)]|0\rangle,
\label{ent-state}
\end{equation}   
where the operator $b_{l\sigma}^{\dagger}(E)$ creates an outgoing
(from the dots towards the beam-splitter) electron plane wave with
spin $\sigma=\uparrow,\downarrow$ and momentum $k(E)=k_F+E/\hbar v_F$
in the normal lead $l=1,2$. Here $k_F$ and $v_F$ is the Fermi wave
number and velocity respectively, same for both normal leads. The
amplitude for this process was found in Ref. [\onlinecite{Recher}] to
have a double-resonant form
\begin{eqnarray}
\langle
  E_1,E_2|T(0)|0\rangle_I&=&\frac{iA_0\gamma/(\pi\sqrt{2})}{(E_1+\varepsilon_1-i\gamma/2)(E_2+\varepsilon_2-i\gamma/2)}.\nonumber
  \\ 
\label{double-res}
\end{eqnarray}
With this we are able to obtain the asymptotics of the outgoing
spin-entangled state. For doing so we substitute Eq.\
(\ref{double-res}) into Eq.\ (\ref{LS-solution2}) and find
\begin{eqnarray}
&&|\Psi_{I}\rangle=|0\rangle + \nonumber \\
&&\int_{-eV}^{eV} dEA(E)[b_{1\uparrow}^{\dagger}(E)b_{2\downarrow}^{\dagger}(-E)
-b_{1\downarrow}^{\dagger}(E)b_{2\uparrow}^{\dagger}(-E)]|0\rangle \nonumber\\
\label{asymptotics1}
\end{eqnarray}   
with
\begin{eqnarray}
A(E)&=&\frac{A_0\gamma}{(E+\varepsilon_1-i\gamma/2)(-E+\varepsilon_2-i\gamma/2)}
\label{double-resx}
\end{eqnarray}
i.e. $A(E)=(-i\pi\sqrt{2})\langle E,-E|T(0)|0\rangle_I$. This state
is the sum of the unperturbed groundstate and an entangled, two
electron state. The entangled state is a linear superposition of spin
singlets at different energies, an entangled two-particle
wavepacket. The singlet spin-state results from the singlet state of
the Cooper-pair, conserved in the tunneling from the
superconductor. Moreover, the two electrons in each singlet have
opposite energies $E$ and $-E$ (counted from $\mu_S=0$), a consequence
of the Cooper-pairs having zero total energy with respect to the
chemical potential of the superconductor.

Several important observations can be made regarding the state in
Eq. (\ref{asymptotics1}). First, the properties, including the
two-particle wavepacket structure, can be clearly seen by writing the
wavefunction in first quantization. Introducing
$|1,E\rangle{\otimes}|\uparrow\rangle$ for
$b_{1}^{\uparrow\dagger}(E)|0\rangle$, the properly symmetrized
wavefunction is given by (omitting the ground state $|0\rangle$)
\begin{eqnarray}
|\Psi_I\rangle&=&\int_{-eV}^{eV}
dEA(E)(|1,E\rangle_{\mu}|2,-E\rangle_{\nu}+|2,-E\rangle_{\mu}|1,E\rangle_{\nu})
\nonumber \\ &&\otimes\left(|\uparrow\rangle_{\mu}
|\downarrow_{\nu}\rangle-|\downarrow\rangle_{\mu} |\uparrow\rangle_{\nu}\right)
\label{state0}
\end{eqnarray}
with $\mu,\nu$ the particle index. The coordinate dependent
wavefunction $\Psi_I(x_{\mu},x_{\nu})=\langle
x_{\mu},x_{\nu}|\Psi_I\rangle$ can then be written ($x=0$ at the
lead-dot connection points)
\begin{eqnarray}
\Psi_I(x_{\mu},x_{\nu})=\psi(x_{\mu},x_{\nu})\left(\lambda_{\mu}^1\lambda_{\nu}^2+\lambda_{\mu}^2\lambda_{\nu}^1
    \right)\left(\chi_{\mu}^{\uparrow}\chi_{\nu}^{\downarrow}-\chi_{\mu}^{\downarrow}\chi_{\nu}^{\uparrow}\right)
    \nonumber \\
\label{state1}
\end{eqnarray}
with 
\begin{eqnarray}
&&\psi(x_{\mu},x_{\nu})=\frac{2\pi i \gamma
    A_0}{2\varepsilon-i\gamma} \nonumber \\
&\times&\mbox{exp}\left[ik_F(x_{\nu}+x_{\mu})-i(\varepsilon-i\gamma/2)|x_{\nu}-x_{\mu}|/\hbar
    v_F\right] \nonumber \\
\end{eqnarray}
where for simplicity the case with energies
$\varepsilon_1=\varepsilon_2\equiv \varepsilon$ is considered. To
arrive at Eq. (\ref{state1}) we first introduced the wavefunctions
$\langle x_l|E \rangle=\mbox{exp}[ik(E)x_l]$, $l=\mu,\nu$, the spin
spinors $\langle x_l|\uparrow\rangle=\chi_l^{\uparrow},\langle
x_l|\downarrow\rangle=\chi_l^{\downarrow}$ and the orbital spinors
$\langle x_l|1\rangle=\lambda_l^{1},\langle
x_l|2\rangle=\lambda_l^{2}$ and then performed the integral over
energy. The orbital spinors describe the wavefunction in the space
formed by the lead indices 1 and 2, a pseudo-spin space, as discussed
in Ref. [\onlinecite{Sam1}]. We note that the beam-splitter, discussed
below, only act in the orbital 12-space (i.e. spin independent
scattering). Moreover, it is the property of the state in 12-space
that determines the current correlators discussed below.

As is clear from Eq. (\ref{state1}), the state is a direct product
state between the spin and orbital part of the wavefunction. The spin
state is antisymmetric under exchange of the two electrons, a singlet,
while the orbital state is symmetric, a triplet. The probability to
jointly detect one electron at $x_{\mu}$ in lead 1 and one at
$x_{\nu}$ in lead 2 decay exponentially with the distance
$|x_{\mu}-x_{\nu}|$, an effect of the two electrons being emitted at
essentially the same time (separated by a small time $\hbar/\Delta$)
to points $x_{\mu}=0$ and $x_{\nu}=0$ respectively. Note that the
state $|\Psi_I\rangle$, a stationary scattering state, does not
describe wave packets in the traditional sense with two electrons
moving out from the dots as time passes (as a solution to the time
independent many particle Schr\"odinger Equation, $|\Psi_I\rangle$ has
a trivial time dependence). To obtain such a wavefunction, one must
break time translation invariance by introducing a time dependent
perturbation, e.g. a variation of the tunnel barrier strength or
dot-level energies in time.

In this context it is worth to mention that such a time dependent
wavefunction was recently considered by Hu and Das Sarma
[\onlinecite{Dassarma}] for a double-dot turnstile entangler. However,
in Ref. [\onlinecite{Dassarma}], the entangled wavefunction was not
derived from a microscopic calculation but merely postulated. The
wavefunction had an amplitude of order unity (no tunneling limit) and
contained a double integral over energy. This is different from our
wavefunction in Eq. (\ref{asymptotics1}) and moreover gives rise to a
qualitatively different result for the currents as well as the current
correlators studied below.

Second, the entangled state in Eq. (\ref{asymptotics1}) has just the
same form as the pair-splitted state obtained in the
normal-superconducting system of Ref. [\onlinecite{Sam1}], where a
scattering approach based on the Bogoliubov-de Gennes equation was
used. This shows rigorously that the effect of the strong Coulomb
blockade, prohibiting two electrons to tunnel through the same dot,
can be incorporated in a scattering formalism by putting the amplitude
for Andreev reflection back into the same dot to zero. From this
observation it follows that the rest of the calculation in the paper
where the state in Eq. (\ref{asymptotics1}) is employed could in
principle be carried out strictly within the scattering approach
\cite{Datta} to the Bogoliubov de Gennes equation. However, in such a
calculation the entanglement is not directly visible, which makes the
interpretation of the result difficult. Instead, below we work
directly with the state in Eq. (\ref{asymptotics1}).

Third, it is also interesting to note the close connection between the
emission of a Cooper pair and the process of spontaneous, parametric
down-conversion\cite{downconv} of pairs of photons investigated in
optics, where a single photon from a pump-laser is split in a
non-linear crystal into two photons. From the point of view of the
theoretical approach, expanding the outgoing state in a ground state
and, to first order in perturbation, an emitted pair of particles, is
similar to the work in e.g. Ref. [\onlinecite{Ou1}]. The resulting
state, Eq. (\ref{asymptotics1}), is a spin singlet, while a state with
polarization entanglement is, under appropriate conditions, produced
in the down-conversion process (type II). Moreover, the emission of
the two electrons is ``spontaneous'', i.e. random and uncorrelated in
time, just in the same way as for the down-converted photons. One can
also point out the maybe less obvious relation that the two electrons
emitted from the superconductor carry information about the phase of
the superconducting condensate, just as the two photons carry
information of the phase of the field of the pump-laser. A coherent
superposition of states of pairs of electrons emitted from different
points of the superconductor, can give rise to observables sensitive
to the difference in superconducting phase between the two emission
points, as was demonstrated in Ref. [\onlinecite{Sam1}]. This has its
analog in the photonic experiment with a single, coherent laser
pumping two separate non-linear crystals, presented in
Ref. [\onlinecite{Ou2}].

\subsection{Electrons tunneling through the same dot.}

We then turn to process II, when the amplitude for tunneling through
the same dot is much larger than the amplitude to tunnel through
different dots. The wavefunction for two electrons
to tunnel to energies $E_1$ and $E_2$ in lead $j$ is
\begin{equation}
|E_1,E_2\rangle_{II}=
\frac{1}{\sqrt{2}}[b_{j\uparrow}^{\dagger}(E_1)b_{j\downarrow}^{\dagger}(E_2)
-b_{j\downarrow}^{\dagger}(E_1)b_{j\uparrow}^{\dagger}(E_2)]|0\rangle,
\label{ent-state2}
\end{equation}   
The amplitude for this process, $\langle E_1,E_2|T(0)|0\rangle_{II}$,
was found in Ref. [\onlinecite{Recher}] to have a single resonant
form, different from Eq. (\ref{double-res}),
\begin{eqnarray}
&&\langle E_1,E_2|T(0)|0\rangle_{II}=\frac{iB_0}{\pi2\sqrt{2}}\nonumber \\
&\times&\left(\frac{1}{E_1+\varepsilon_j-i\gamma/2}+\frac{1}{E_2+\varepsilon_j-i\gamma/2}\right).
\label{double-res2}
\end{eqnarray}
Here, for simplicity the two dot-superconductor contacts are taken to
be identical. Since the superconductor is a macroscopically coherent
object, the total state is a linear combination of the states
corresponding to two electrons tunneling through dot 1 and dot 2. To
obtain the asymptotics of the outgoing spin-entangled state, we
substitute Eq.\ (\ref{double-res2}) into Eq.\ (\ref{LS-solution2}) and
find
\begin{eqnarray}
&&|\Psi_{II}\rangle=|0\rangle+\int_{-eV}^{eV}dE\left[B_{1}(E)b_{1\uparrow}^{\dagger}(E)b_{1\downarrow}^{\dagger}(-E)\right.\nonumber \\
&&\left. +B_2(E)b_{2\uparrow}^{\dagger}(E)b_{2\downarrow}^{\dagger}(-E) \right]  
\label{asymptotics3}
\end{eqnarray} 
with
\begin{eqnarray}
B_j(E)&=&\frac{B_0(\varepsilon_j-i\gamma/2)}{(E+\varepsilon_j-i\gamma/2)(-E+\varepsilon_j-i\gamma/2)}
\label{double-res2b}
\end{eqnarray}
i.e. $B_j(E)=(-i\pi2\sqrt{2})\langle E,-E|T(0)|0\rangle_{II}$. Arriving
  at Eq. (\ref{asymptotics3}) we used the property $B(-E)=B(E)$ and
  the anti-commutation relations of the fermionic operators.

This state is a linear superposition of the states for two electrons
tunneling through the same dot. Comparing to the state
$|\Psi_I\rangle$ in Eq. (\ref{state1}) for the two electrons tunneling
through different dots, we can make the following comments: (i) Just
as $|\Psi_I\rangle$, the wavefunction $|\Psi_{II}\rangle$ in first
quantization is a product of a an orbital and a spin wavefunction. The
spin wavefunction is, as for $|\Psi_I\rangle$, a singlet
$\chi_{\mu}^{\uparrow}\chi_{\nu}^{\downarrow}-\chi_{\mu}^{\downarrow}\chi_{\nu}^{\uparrow}$.
The orbital wavefunction for the simplest situation
$\varepsilon_1=\varepsilon_2$ is however proportional to
$\lambda_{\mu}^1\lambda_{\nu}^1+\lambda_{\mu}^2\lambda_{\nu}^2$, one
of the Bell states, an orbitally entangled state. (ii) The state
$|\Psi_{II}\rangle$ is the same that would be obtained within
scattering theory (as was shown in Ref. [\onlinecite{Sam1}]), taking
$B_j(E)$ to be the effective Andreev reflection amplitude at dot $j$
and assuming no crossed Andreev reflection between the dots, i.e. zero
probability for an incident electron in lead 1 to be back-reflected as
a hole in lead 2 and vice versa.

With the state in Eq. (\ref{asymptotics3}) and the state for two
electrons tunneling through different dots, in
Eq. (\ref{asymptotics1}), we are in a position to analyze the
transport properties.

\section{Current correlators} 

The two electrons emitted from the dot-superconductor entangler
propagate in the leads $1$ and $2$ towards the normal reservoirs $A$
and $B$. As shown in Fig. \ref{fig1}, the two normal leads are crossed
in a single mode reflectionless beam-splitter. The beam-splitter is
characterized by a spin- and energy independent unitary scattering
matrix connecting outgoing and ingoing operators as
\begin{equation}
\left(\begin{array}{c} b_{A} \\ b_{B}\end{array}
\right)=\left(\begin{array}{cc} r & t' \\ t & r' \end{array}
\right)\left(\begin{array}{c} b_{1} \\ b_{2}\end{array} \right),
\label{smat}
\end{equation}   
where the subscript $\alpha=A,B$ denotes towards what reservoir the
electron is propagating.  The electrons are then detected in the
normal reservoirs $A$ and $B$. 

We point out that beam-splitters without backscattering are not easily
produced experimentally, thus in a more detailed model one should also
take the effect of back scattering into account. Several aspects of
back scattering were recently investigated by Burkard and Loss
,\cite{Burk03} extending the model in
Ref. [\onlinecite{Burk00}]. Although back-scattering can be
incorporated in our model as well, this would complicate the
calculations and make the result less transparent. Below, we instead
neglect the effect of backscattering (however, some qualitative
aspects are discussed below, in the context of the Aharonov-Bohm
effect), pointing out that result below holds rigorously only in the
case with negligible back-scattering at the beam-splitter. 

We also note that in the typical system of interest, with a lateral
size $L$ in the micrometer range, the energy-dependent part of the
phase $\sim L\gamma/\hbar v_F$ picked up by the electrons when
propagating in the leads is negligiably small. Energy independent
phases due to propagation can be included in the scattering amplitudes
of the beam-splitter.

The properties of the electrons emitted by the entangler are
investigated via the current and the zero-frequency current
correlators. The electrical current operator in lead $\alpha$ is given
by \cite{butt90}
\begin{eqnarray}
\hat I_{\alpha}&=&\frac{e}{h}\int dEdE' e^{i(E-E')t/\hbar} \nonumber \\
&\times&\sum_{\sigma}
\left[b^{\dagger}_{\alpha\sigma}(E)b_{\alpha\sigma}(E')-a^{\dagger}_{\alpha\sigma}(E)a_{\alpha\sigma}(E')\right],
\end{eqnarray}   
where $a^{\dagger}_{\alpha\sigma}(E)$ creates an electron plane wave
incoming from the normal reservoir $\alpha$ with spin
$\sigma=\uparrow,\downarrow$ and momentum $k(E)$. The averaged current
is given by
\begin{eqnarray}
I_{\alpha}\equiv\langle \hat I_{\alpha} \rangle,
\label{avcurr}
\end{eqnarray}  
where $\langle ... \rangle\equiv \langle \Psi|...| \Psi \rangle$. The
 zero-frequency correlations between the currents in the leads
 $\alpha$ and $\beta$ are
\begin{equation}
S_{\alpha\beta}=\int dt \langle \Delta \hat I_{\alpha}(t)\Delta \hat
I_{\beta}(0)+\Delta \hat I_{\beta}(0)\Delta \hat I_{\alpha}(t)\rangle,
\label{noise}
\end{equation}   
where $\Delta I_{\alpha}(t)=I_{\alpha}(t)-I_{\alpha}$ is the
fluctuating part of the current in lead $\alpha$. We study the two
cases with electrons tunneling through different dots and the same dot
separately.

\section{Tunneling through different dots.}

For electrons tunneling through different dots, the question is how
the degree of spin-singlet entanglement is reflected in the current
and current correlators. The averaged current, evaluated with the
state $|\Psi_I\rangle$ in Eq. (\ref{asymptotics1}), becomes
\begin{equation}
I_{\alpha}^I=\frac{2e}{h} \int_{-eV}^{eV} dE |A(E)|^2, 
\end{equation}   
same for both $\alpha=A$ and $B$.  Since the two resonances
$\varepsilon_1$ and $\varepsilon_2$ are well within the voltage range,
i.e. $eV-|\varepsilon_1|,eV-|\varepsilon_2|\gg \gamma$, we get the
current
\begin{equation}
I_{\alpha}^I=\frac{2e}{h}\frac{4\pi|A_0|^2\gamma}{(\epsilon_1+\epsilon_2)^2+\gamma^2},
\label{currexp}
\end{equation}   
just the same expression as in Ref. [\onlinecite{Recher}], where the
leads of the entangler were contacted directly to the normal
reservoirs (no beam-splitter). The current is maximal for an asymmetric
setting of the resonances $\varepsilon_1=-\varepsilon_2$. This
two-particle resonance reflects the fact that the two electrons in the
Cooper pairs are emitted at opposite energies with respect to the
superconducting chemical potential. The current contains no
information about the entanglement of the emitted state. In fact, the
same current would be obtained by considering a product state of one
electron in lead 1 and one in lead 2, independent of their spins.

To obtain information about the entanglement, we turn to the current
correlators. Inserting the expression for the state $|\Psi_I\rangle$
into Eq. (\ref{noise}), following Ref. [\onlinecite{butt90}], we get
the expressions for the auto-correlations
\begin{eqnarray}
S_{AA}^{I}=S_{BB}^{I}&=&\frac{4e^2}{h}\int_{-eV}^{eV} dE \left\{[1+2RT]|A(E)|^2 \right. \nonumber\\
&&\left. +2RTA(E)A^*(-E)\right\}
\label{autonoise}
\end{eqnarray}   
as well as the cross-correlations
\begin{eqnarray}
S_{AB}^{I}=S_{BA}^{I}&=&\frac{4e^2}{h}\int_{-eV}^{eV} dE\left\{[T^2+R^2]|A(E)|^2 \right.
\nonumber \\
 &&\left. -2RTA(E)A^*(-E) \right\},
\label{crossnoise}
\end{eqnarray}   
where $R=|r|^2=|r'|^2$ and $T=|t|^2=|t'|^2=1-R$. We note that the
total noise $S^I$ of the current flowing out of the superconductor
is twice the Poissonian, i.e.
\begin{equation}
S^I=S_{AB}^I+S_{BA}^I+S_{AA}^I+S_{BB}^I=4e(I_A^I+I_B^I),
\end{equation}   
describing an uncorrelated emission of pairs of electrons. This
result, an effect of the tunneling limit, is different from the one in
Ref. [\onlinecite{Burk00}] where an entangled state with unity
amplitude was considered and the total noise was found to be zero.

It is clear from the calculation that the second term in
Eqs. (\ref{autonoise}) and (\ref{crossnoise}) depends directly on the
symmetry properties of the orbital wavefunction, and thus, due to the
anti-symmetry of the total wavefunction, indirectly on the symmetry
properties of the spin wavefunction. For a spin-triplet state
$|\Psi_I\rangle$ the last term in Eqs. (\ref{autonoise}) and
(\ref{crossnoise}) would have opposite sign. Since all the three
possible triplets, with spin wavefunctions
$\chi_{\mu}^{\uparrow}\chi_{\nu}^{\downarrow}+\chi_{\mu}^{\downarrow}\chi_{\nu}^{\uparrow},~\chi_{\mu}^{\uparrow}\chi_{\nu}^{\uparrow}$
and $\chi_{\mu}^{\downarrow}\chi_{\nu}^{\downarrow}$ have the same
anti-symmetric orbital wavefunction
($\lambda_{\mu}^1\lambda_{\nu}^2-\lambda_{\mu}^2\lambda_{\nu}^1$ for
$\varepsilon_1=\varepsilon_2$) they give rise to the same noise
correlators. As a consequence, performing a noise correlation
measurement, one can only distinghuish between spin-singlets and
spin-triplets, but not between entangled
$\chi_{\mu}^{\uparrow}\chi_{\nu}^{\downarrow}+\chi_{\mu}^{\downarrow}\chi_{\nu}^{\uparrow}$
and non-entangled
$\chi_{\mu}^{\uparrow}\chi_{\nu}^{\uparrow},~\chi_{\mu}^{\downarrow}\chi_{\nu}^{\downarrow}$
spin-triplets. This was pointed out already in
Ref. [\onlinecite{Burk00}]. We note that it is possible to
distinguish between the different triplets in a more advanced
beamsplitter scheme, using controlled single spin rotations via a
e.g. local Rashba interaction.\cite{Egues} Such a scheme is
straightforwardly included into our theoretical treatment, however, it
demands a more involved experimental setup and is therefore not
considered here, we restrict our investigation to the simplest
possible system.

To investigate the properties of the current correlators in detail,
the remaining integral over energy in Eqs. (\ref{autonoise}) and
(\ref{crossnoise}) is carried out, giving
\begin{equation}
\int dE
A(E)A^*(-E)=\frac{4\pi|A_0|^2\gamma^3}{[(\epsilon_1-\epsilon_2)^2+\gamma^2][(\epsilon_1+\epsilon_2)^2+\gamma^2]}.
\label{integral}
\end{equation}   
This shows that, unlike the current, the noise is sensitive to both
the difference and the sum of the dot energy levels. We note that the
integral of $A(E)A^*(-E)$ is manifestly positive and smaller than the
integral of $|A(E)|^2$ for all $\varepsilon_1,\varepsilon_2$ except
for $\varepsilon_1=\varepsilon_2$, when they are equal.

From these observations we can draw several conclusions and compare
our results to the results in Ref. [\onlinecite{Burk00}]:

$(i)$ The second term in Eqs. (\ref{autonoise}) and
(\ref{crossnoise}), dependent on the orbital symmetry of the
wavefunction, leads to a suppression of the cross-correlation but an
enhancement of the auto-correlation. This is an effect of the bunching
behavior of the spin-singlet, i.e. the two electrons show an increased
probability to end up in the same normal reservoir. \cite{Burk00} For
a symmetric beam-splitter, $R=T=1/2$ and aligned dot-levels
$\varepsilon_1=\varepsilon_2$, the cross-correlations are zero (to the
leading order in tunneling probability considered here). This is a
signature of perfect bunching of the two electrons.

$(ii)$ The last term in Eqs. (\ref{autonoise}) and (\ref{crossnoise})
is proportional the spectral overlap $\int dE A(E)A^*(-E)$. The
spectral overlap physically corresponds to the overlap between the
wavefunctions of the two electrons colliding in the beam-splitter. For
single-particle levels at different energies $\varepsilon_1 \neq
\varepsilon_2$, the spectral amplitudes of the emitted electrons are
centered at different energies and consequently\cite{Burk00} the Pauli
principle responsible for the bunching is less efficient.

It is important to note that the last term in Eqs. (\ref{autonoise})
and (\ref{crossnoise}), dependent on the bunching, generally is of the
same magnitude as the first term. We emphasize that this result is
qualitatively different from what was found in
Ref. [\onlinecite{Burk00}], where the bunching dependent part of the
current correlator was proportional to a Kronecker delta-function in
energy, a consequence of considering a discrete spectrum. Our result
clearly shows that it should be experimentally feasible to detect the
bunching, and thus demonstrate that spin singlets are emitted from the
entangler. We note that the same qualitative result was found in
Ref. [\onlinecite{Dassarma}].

$(iii)$ The cross correlations are positive for any transparency of
the beam-splitters (note that $R^2+T^2 \geq 2RT$). This is different
from the result in Ref. [\onlinecite{Burk00}], where negative cross
correlations were predicted. The negative correlations are again a
result of the unity amplitude of the incoming entangled state
considered in Ref. [\onlinecite{Burk00}]. In this context, we point out
that positive cross-correlations have been predicted in several few
mode \cite{poscorrNS1} and many mode \cite{poscorrNS2}
normal-superconductors hybrid systems as well as purely normal systems
in the presence of interactions.\cite{poscorrNN1} In several of these
cases, the positive correlations were explained with semiclassical
models. Thus, the presence of positive correlations can not itself be
taken as a sign of spin-entanglement.

\begin{figure}[h]
\centerline{\psfig{figure=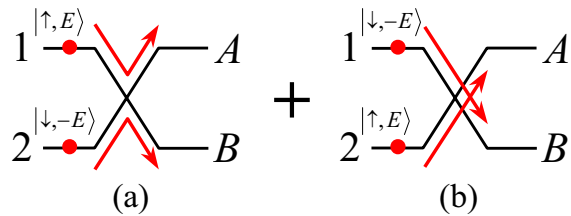,width=0.9\linewidth}}
\caption{Elementary scattering processes (shown at the beamsplitter)
contributing to the cross correlators $S_{AB}^I$. The two processes
(a) and (b) transport a pair of electrons $|\uparrow,E\rangle$ and
$|\downarrow,-E\rangle$ from the superconductor to the reservoirs $A$
and $B$ respectively. The two processes, having the same initial and
final state, are indistinguishable and their amplitudes must be
added. The correlator $S_{AB}^I$ is proportional to the integral over
energy of the (energy dependent) joint detection probability.}
\label{figAB}
\end{figure}

We point out that the expression for the energy dependent integrand of
the cross correlators in Eq. (\ref{crossnoise}) can be understood in
an intuitive way, by considering the elementary scattering processes
contributing to the noise, shown in Fig. \ref{figAB}.
Let us consider the probability for the two electrons emitted from the
superconductor to end up, one with spin up and energy $E$ in reservoir
$A$ and the other with spin down and energy $-E$ in reservoir
$B$. There are two paths the electrons can take from the
superconductor to the reservoirs: (a) the electron with spin up and
energy $E$ via dot 1 and the electron with spin down and energy $-E$
via dot 2. This process has an amplitude $tt'A(E)$ (b) the electron
with spin down and energy $-E$ via dot 1 and the electron with spin
down and energy $-E$ via dot 2. This process has an amplitude
$rr'A(-E)$. Since the two processes have the same initial and final
states, they are indistinguishable and their amplitudes must be added.
This gives together the energy dependent joint detection probability
$\sim
|tt'A(E)+rr'A(-E)|^2=T^2|A(E)|+R^2|A(-E)|^2+rr't^*t^{'*}A(E)A^*(-E)+r^*r^{'*}tt'A(-E)A^*(E)$.
In analogy to the noise correlators for the entangler with energy
independent tunneling probabilities in Ref. [\onlinecite{Sam1}], it is
found that the noise correlator $S_{AB}^I$ is simply proportional to
integral over energy of the joint detection probability. Using that
the integral in Eq. (\ref{crossnoise}) goes from $-eV$ to $eV$ and
that the unitarity of the scattering matrix in Eq. (\ref{smat}) gives
$rt^*+t'r^{'*}=0$, we get the expression in the integrand in
Eq. (\ref{crossnoise}).

For the auto-correlation, a similar interpretation in terms of
probabilities for two-particle scattering processes only is not
possible, one also has to consider single particle
probabilities. Formally, this is the case since auto-correlations
contain exchange effects between the two particles scattering to the
same reservoir.

\subsection{Fano factors}
\begin{figure}[h]
\centerline{\psfig{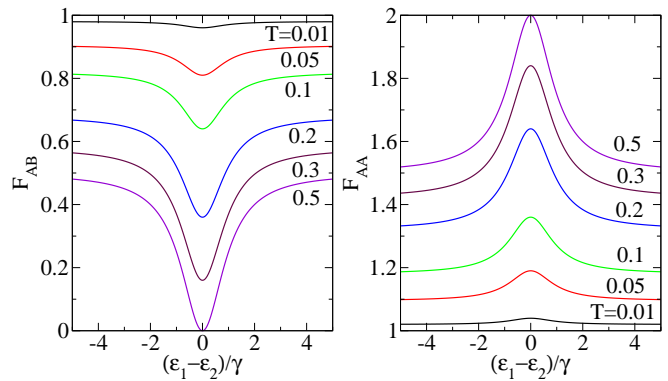}}
\caption{The Fano factor for the cross-correlations $F_{AB}=F_{BA}$
  (left panel) and auto-correlations $F_{AA}=F_{BB}$ (right panel) as
  a function of the normalized energy difference
  $(\varepsilon_1-\varepsilon_2)/\gamma$ for various beam-splitter
  transparencies.}
\label{fig2}
\end{figure}

A quantitative analysis of the current correlators is most naturally
performed via the Fano factors
$F_{\alpha\beta}=S_{\alpha\beta}/(2e\sqrt{I_{\alpha}I_{\beta}})$. The
Fano factor isolates the dependence of the noise on various
parameters, not already present in the current. For the cross- and
auto-correlations respectively, we have
\begin{eqnarray}
F_{AB}^{I}=F_{BA}^I=T^2+R^2-2RT|H(\varepsilon_1-\varepsilon_2)|^2
\label{crossfano}
\end{eqnarray}   
and
\begin{eqnarray}
F_{AA}^{I}=F_{BB}^I=1+2RT+2RT|H(\varepsilon_1-\varepsilon_2)|^2
\label{autofano}
\end{eqnarray}   
where
\begin{eqnarray}
H(\varepsilon_1-\varepsilon_2)=\frac{i\gamma}{\varepsilon_1-\varepsilon_2+i\gamma}.
\label{hexpr}
\end{eqnarray}   
We note that only the last terms in Eqs. (\ref{crossfano}) and
(\ref{autofano}) depend on the energies $\varepsilon_1$ and
$\varepsilon_2$ of the levels in the dots. The Fano factor as a
function of energy difference $\varepsilon_1-\varepsilon_2$ is plotted
for several values of transparency of the beam splitter in
Fig. \ref{fig2}.
For the cross-correlators, the Fano factor has a minimum for the two
resonant levels aligned, $\varepsilon_1=\varepsilon_2$. The value at
this minimum decreases monotonically from $1$ to $0$ when increasing
the transparency $T$ of the beam-splitters from $0$ to $0.5$ (the Fano
factor for transmission probability $T$ is the same as for
$1-T$). Thus, for a completely symmetric beam-splitter, $T=R=0.5$, the
Fano factor is zero. This corresponds to the case of perfect
bunching. For the auto-correlators, the picture is the opposite. The
Fano factor has a maximum for the two resonances aligned,
$\varepsilon_1=\varepsilon_2$. The value at this maximum increases
monotonically from $1$ to $2$ when increasing the transparency $T$ of
the beam-splitters from $0$ to $0.5$. Thus, for a symmetric
beam-splitter, $T=R=0.5$, the Fano factor is now two.

\subsection{Decoherence}

Considering the robustness of the bunching behavior, an important
observation is that the Fano factors in Eqs. (\ref{autofano}) and
(\ref{crossfano}) [as well as the noise correlators in
Eqs. (\ref{autonoise}) and (\ref{crossnoise})] only depend on the
transmission and reflection probabilities $T$ and $R$. All information
about the scattering phases, from the beam-splitter as well as from
the propagation in the leads, drops out. As a consequence, the
correlators are insensitive to dephasing of the orbital part of the
wavefunction, i.e. processes that cause slow and energy independent
fluctuations of the scattering phases. This insensitivity, different
from schemes based on orbital entanglement,
\cite{Sam1,Been1,samHBT,dephase} can be understood by considering the
first quantized version [in Eq. (\ref{state1})] of the wavefunction
$|\Psi_I\rangle$. Any orbital phase picked up by an electron in
e.g. lead 1 just gives rise to an overall phasefactor of the total
orbital wavefunction, since each term in the wavefunction corresponds
to one electron in lead 1 and one in lead 2. Moreover, any orbital
``pseudo spin flip'' would imply a scattering of particles between the
leads 1 and 2 and is not allowed in the non-local geometry.

The situation is different for spin decoherence, energy independent
spin-flip or spin-dephasing processes tending to randomize the spin
directions. Spin decoherence generally modifies the Fano factors in
Eqs. (\ref{crossfano}) and (\ref{autofano}). Formally, the (mixed)
state in the presence of decoherence is described by a density matrix
$\rho$. Writing $\rho$ in a spin singlet-triplet basis, as shown in
the Appendix, only the diagonal elements $\rho_{SS}$ (singlet) and
$\rho_{T_0T_0}, \rho_{T_+T_+},\rho_{T_-T_-}$ (triplets) contribute to
the current correlators. As discussed above, all the three spin
triplets give rise to the same Fano factors. The spin-triplet Fano
factors are given by the spin-singlet ones in Eqs. (\ref{crossfano})
and (\ref{autofano}) by changing the sign of the last term
$2RT|H(\varepsilon_1-\varepsilon_2)|^2$, i.e. from bunching to
anti-bunching. Using that the sum of the diagonal elements of the
density matrix is one,
i.e. $\rho_{SS}+\rho_{T_0T_0}+\rho_{T_+T_+}+\rho_{T_-T_-}=1$, the
effect of spin decoherence is to renormalize only the part of the
Fano factors dependent on the dot-level energies as
\begin{equation}
|H(\varepsilon_1-\varepsilon_2)|^2\rightarrow(2\rho_{SS}-1)|H(\varepsilon_1-\varepsilon_2)|^2.
\label{renormeq}
\end{equation}
The renormalization factor is thus the singlet weight minus the total
triplet weight,
$\rho_{SS}-(\rho_{T_0T_0}+\rho_{T_+T_+}+\rho_{T_-T_-})=2\rho_{SS}-1$.
This clearly displays how decoherence, reducing the singlet weight and
consequently increasing the triplet weight, leads to a cross-over at
$\rho_{SS}=1/2$ from a bunching to an anti-bunching behavior of the
noise correlators. For a completely dephased spin state, with an equal
mixture of singlets and triplets
($\rho_{SS}=\rho_{T_0T_0}=\rho_{T_+T_+}=\rho_{T_-T_-}=1/4$), the
renormalization factor $2\rho_{SS}-1$ saturates at the value $-1/2$.

We point out that this discussion might be modified when considering
other types of effects causing decoherence, such as e.g. inelastic
scattering. A more detailed investigation (see
e.g. Refs. [\onlinecite{Marq}]), going beyond the scope of the paper,
is needed to address these issues.

\subsection{Spin entanglement bound}

In the absence of spin decoherence the spin state of the emitted pair
is a singlet, a maximally entangled state. For finite spin
decoherence, this is no longer the case and the question arises how to
obtain quantitative information about the spin entanglement from the
measurements of the current correlators.

We stress that our interest here is the spin entanglement of $\rho$
only. However, $\rho$ contains information about the spin part of the
state as well as the energy-dependent orbital part, the wave-packet
structure of the emitted pair of electrons. To quantify the spin
entanglement, one thus has to consider a measurement sensitive to the
spin part of $\rho$ only (see Appendix). One such important example is
the cross correlators between the currents in the leads 1 and 2
(i.e. without beamsplitters). It was shown in a related system in
Ref. [\onlinecite{Sam1}] that these cross correlators are simply
proportional to the probability to jointly detect one particle in lead
1 and one in 2.  The wave-packet property of the emitted pair results
only in an overall constant multiplying the probabilities. As a
consequence, a Bell Inequality, derived in terms of the joint
detection probabilities, could be formulated in terms of
zero-frequency cross correlators. In the same way, for the
superconductor-dot entangler considered here, the spin entanglement of
the two emitted electrons can in principle be tested via a Bell
Inequality formulated in terms spin current correlators. \cite{BI} The
situation is different for the beam-splitter setup, where the Fano
factors in Eqs. (\ref{crossfano}) and (\ref{autofano}) in general
depend on the wave-packet structure via the dot-level dependent factor
$|H(\varepsilon_1-\varepsilon_2)|^2$, quantifying the overlap of the
two electrons when colliding. However, for
$\varepsilon_1=\varepsilon_2$ (i.e. maximal overlap, $|H(0)|^2=1$) the
Fano factors are independent of the wave-packet structure of the
emitted electrons and thus only sensitive to the spin part of $\rho$
[see Eq. (\ref{renormeq})].

The spin part of $\rho$ can be described by the $4\times 4$ spin
density matrix $\rho_{\sigma}$, rigorously defined in the Appendix
(note that $\rho$, due to the continous energy variable, is infinite
dimensional). Formally, $\rho_{\sigma}$ is the density matrix obtained
when tracing $\rho$, for aligned dot-levels
$\varepsilon_1=\varepsilon_2$, over energies. The question is thus how
to determine the entanglement of $\rho_{\sigma}$. In general,
knowledge of all the matrix elements is needed. This information can
however not be obtained within our approach, since the Fano factors
only provides information of the spin singlet weight, as is clear from
Eq. (\ref{renormeq}). It is nevertheless possible, as described in
detail in the Appendix, to follow the ideas of Burkard and Loss
\cite{Burk03} and obtain a lower bound for the spin entanglement.

There are several different measures of entanglement for the mixed
state of two coupled spin-$1/2$ systems. Here we consider the
concurrence\cite{Wootters} $C$, with $C=0$ $(C=1)$ for an unentangled
(maximally entangled) state. To establish the lower bound, it can be
shown that the concurrence $C(\rho_{\sigma})$ is always larger than or
equal to the concurrence $C(\rho_W)$ of the so Werner
state,\cite{Werner} described by the density matrix $\rho_W$. The
Werner state, defined as the average of $\rho_{\sigma}$ over identical
and local random rotations, has the same singlet weight $\rho_{SS}$ as
$\rho_{\sigma}$. The concurrence of the Werner state has the appealing
property that it is a function of the spin singlet weight only,
$C_W=\mbox{max}\{2\rho_{SS}-1,0\}$.

The findings above thus lead to the simple and important result that
the renormalization, Eq. (\ref{renormeq}), of the Fano factors in
Eqs. (\ref{crossfano}) and (\ref{autofano}) due to spin decoherence
can be written as (for $C_W>0$)
\begin{eqnarray}
|H(\varepsilon_1-\varepsilon_2)|^2\rightarrow C_{W}|H(\varepsilon_1-\varepsilon_2)|^2 
\label{renorm}
\end{eqnarray}   
where $C_W$ thus provides a lower bound for the spin entanglement of
the emitted pair of electrons (for the pure singlet $\rho_{SS}=1$,
$C_W$ and $C(\rho_{\sigma})$ are equal and maximal). Thus, as long as
the Fano factors display a bunching behavior, the spin entanglement is
finite, $C_{W}>0$. For a cross-over to anti-bunching behavior,
$C_{W}=0$ and one can no longer conclude anything about the
entanglement of the spins state.  The value of $C_{W}$ can be
extracted directly from the experimentally determined Fano factors, as
the amplitude of the modulation of the Fano factors with respect to
dot level amplitudes $\varepsilon_1-\varepsilon_2$ divided by
$2RT$. The values of $R,T$ can be extracted independently from the
Fano factors at dot levels such that
$H(\varepsilon_1-\varepsilon_2)\approx 0$.

The result in Eq. (\ref{renorm}) thus provides a simple relation
between the Fano-factors and the minimum spin entanglement $C_{W}$. It
is clear, however, that since the Fano factors only provide
information of the singlet weight, full information of the spin
entanglement can not be obtained by the beam-splitter approach
employed here. It should be noted that the result in
Eq. (\ref{renorm}) is quantitatively different from what was obtained
in Ref. [\onlinecite{Burk03}], a consequence of the different states
considered for the emitted electrons, as discussed above in connection
with the current correlators.

\section{Tunneling through the same dot.}

We then turn to the situation when the two electrons tunnel through
the same dot. To be able to distinguish this process II from process
I, it is important to study the current as well as the noise in
detail. The averaged current in Eq. (\ref{avcurr}), evaluated with the
state in Eq. (\ref{asymptotics3}), becomes for reservoirs $A$ and $B$
\begin{eqnarray}
I_{A}^{II}=\frac{2e}{h}\int_{-eV}^{eV} dE
\left(R|B_1{}(E)|^2+T|B_{2}(E)|^2\right), \nonumber \\
I_{B}^{II}=\frac{2e}{h}\int_{-eV}^{eV} dE
\left(T|B_1{}(E)|^2+R|B_{2}(E)|^2\right).
\end{eqnarray}   
Since the two resonances $\varepsilon_1$ and $\varepsilon_2$ are well
within the voltage range,
i.e. $eV-|\varepsilon_1|$, $eV-|\varepsilon_2|\gg \gamma$, we can
perform the integrals and get the current \cite{Recher}
\begin{equation}
I^{II}_{\alpha}=\frac{2e}{h}\pi|B_0|^2/\gamma,
\label{currexp2}
\end{equation}   
the same for both reservoirs $\alpha=A,B$. We note that the
two-particle resonance in the current, present in the pair-splitting
case I, is absent due to the Coulomb blockade, as pointed out in
Ref. [\onlinecite{Recher}]. A difference from
Ref. [\onlinecite{Recher}] is however that due to the absence of
back-scattering at the beam-splitter, there is no scattering-phase
dependence of the current. Consequently, there is no dependence on a
possible difference in the superconducting phase at the two emission
points or an Aharonov-Bohm phase \cite{AB} due to a magnetic flux in
the area between the superconductor, the dots and the
beam-splitter. It should be pointed out that this is not a generic
result for normal-superconducting systems. In a situation with
backscattering, which is inevitable in e.g. the three-terminal
fork-like geometries, Andreev interferometers, studied extensively in
both diffusive\cite{NazAI} and ballistic\cite{ballAI} conductors, the
current is indeed sensitive to a superconducting phase difference as
well as the Aharonov-Bohm phase.

Regarding the spin entanglement, just as for process I, no information
is provided by the averaged current. The same result would have been
obtained considering an incoherent superposition of two electrons in
lead 1 and two in lead 2, independent on spin state. Turning to the
current correlators, inserting the expression for the state
$|\Psi_{II}\rangle$ into Eq. (\ref{noise}), one gets the expressions
for the auto-correlations
\begin{eqnarray}
&&S_{AA}^{II}=\frac{4e^2}{h}\int dE \left\{R(1+R)|B_1(E)|^2 \right. \nonumber\\
&&\left.+T(1+T)|B_2(E)|^2+2
    \mbox{Re}\left[(r^*t')^2B_1^*(E)B_2(E)\right]\right\},\nonumber \\
&&S_{BB}^{II}=\frac{4e^2}{h}\int dE \left\{T(1+T)|B_1(E)|^2 \right. \nonumber\\
&&\left.+R(1+R)|B_2(E)|^2+2
    \mbox{Re}\left[(r^*t')^2B_1^*(E)B_2(E)\right]\right\}\nonumber \\
\label{autonoise2}
\end{eqnarray}   
with Re$[..]$ denoting the real part, as well as the cross-correlations
\begin{eqnarray}
S_{AB}^{II}=S_{BA}^{II}&=&\frac{4e^2}{h}\int dE \left\{RT(|B_1(E)|^2+|B_2(E)|^2) \right. \nonumber\\
&&\left.-2
    \mbox{Re}\left[(r^*t')^2B_1^*(E)B_2(E)\right]\right\}.
\label{crossnoise2}
\end{eqnarray}   
The integrals over $|B_j(E)|^2$ were carried out above,
Eq. (\ref{currexp2}). Performing the integral over $B_1(E)B_2^*(-E)$
in the limit $eV-|\varepsilon_1|$,$eV-|\varepsilon_2|\gg \gamma$, we
get
\begin{equation}
\int dE B_1^*(E)B_2(E)=\frac{\pi i |B_0|^2}{\epsilon_1-\epsilon_2+i\gamma}.
\label{integral2}
\end{equation}   
The expressions for the correlators above give that the total noise
$S^{II}$ of the current flowing out of the superconductors is,
\begin{equation}
S^{II}=S_{AB}^{II}+S_{BA}^{II}+S_{AA}^{II}+S_{BB}^{II}=4e(I_A^{II}+I_B^{II}),
\end{equation}   
twice the Poissonian, describing, just as in case I, an uncorrelated
emission of pairs of electrons.

We note that in contrast to the current and the transport properties
in case I, when the two electrons tunnel through different dots, the
noise contains information about the scattering phases (via
$r^*t'$). Quite generally, one can write
\begin{equation}
(r^*t')^2=RTe^{i\phi}, 
\end{equation}   
where $\phi$ is a scattering phase of the beam-splitter. Scattering
phases picked up during propagation in the leads simply add to
$\phi$. As a consequence, $\phi$ can be modulated by e.g. an
electrostatic gate changing the length of the lead $1$ or $2$ or by an
Aharonov-Bohm flux threading the region between the dots, the
superconductor and the beam-splitter. An important consequence of this
phase dependence of the current correlators is that it can be used to
distinguish between tunneling via process II and between process I,
since the current correlators of the latter show no phase
dependence. This was pointed out in Ref. [\onlinecite{Recher}].

This phase dependence shows that the correlators in
Eqs. (\ref{autonoise2}) and (\ref{crossnoise2}) are sensitive to
dephasing affecting the orbital part of the wavefunction. For complete
dephasing, the last term in Eqs. (\ref{autonoise2}) and
(\ref{crossnoise2}) are suppressed. The orbital entanglement in
Eq. (\ref{asymptotics3}), the linear superposition of states
corresponding to tunneling through dot 1 and 2, is lost. This
sensitivity to orbital dephasing is different from the one for process
I discussed above. However, again in contrast to process I, the
current correlators are insensitive to spin-dephasing. This can be
understood by considering the first quantized wavefunction
$|\Psi_{II}\rangle$, discussed below Eq. (\ref{double-res2b}), keeping
in mind that the wavefunction is a direct product of a spin part and
an orbital part. The spin wavefunction is a singlet,
$\chi_{\mu}^{\uparrow}\chi_{\nu}^{\downarrow}-\chi^{\downarrow}_{\mu}\chi_{\nu}^{\uparrow}$,
but the orbital wavefunction is a combination of triplets,
$\lambda_{\mu}^1\lambda_{\nu}^1+\lambda_{\mu}^2\lambda_{\nu}^2$ for
$\varepsilon_1=\varepsilon_2$. Since no scattering between the leads
is possible. i.e. no ``pseudo spin flip'', orbital dephasing can not
change the triplet character of the orbital wavefunction and as a
result, the spin wavefunction is bound to be a singlet. Thus, the spin
entanglement in $|\Psi_{II}\rangle $ is protected against decoherence.

Turning to the Fano factor gives, the auto and cross correlations are
\begin{eqnarray}
F_{AA}^{II}=F_{BB}^{II}&=&1+T^2+R^2 \nonumber \\
&+&2RT\mbox{Re}\left[e^{i\phi}H(\varepsilon_1-\varepsilon_2)\right]
\label{crossfano2}
\end{eqnarray}   
and
\begin{eqnarray}
F_{AB}^{II}=F_{BA}^{II}&=&2RT \nonumber \\
&-&2RT\mbox{Re}\left[e^{i\phi}H(\varepsilon_1-\varepsilon_2)\right]
\label{autofano2}
\end{eqnarray}   
respectively, where $H(\varepsilon_1-\varepsilon_2)$ is given in
Eq. (\ref{hexpr}).
\begin{figure}[h]
\centerline{\psfig{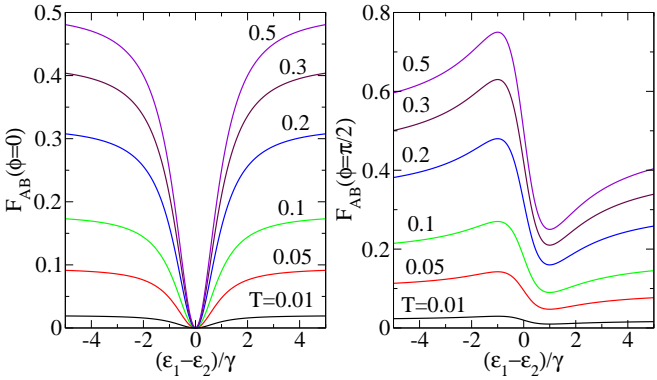}}
\caption{The Fano factor for the cross-correlations $F_{AB}=F_{BA}$ for
  phase difference $\phi=0$ (left panel) $\phi=\pi/2$ (right panel)
  and as a function of the normalized energy difference
  $(\varepsilon_1-\varepsilon_2)/\gamma$ for various beam-splitter
  transparencies.} 
\label{fig4}
\end{figure}

\begin{figure}[h]
\centerline{\psfig{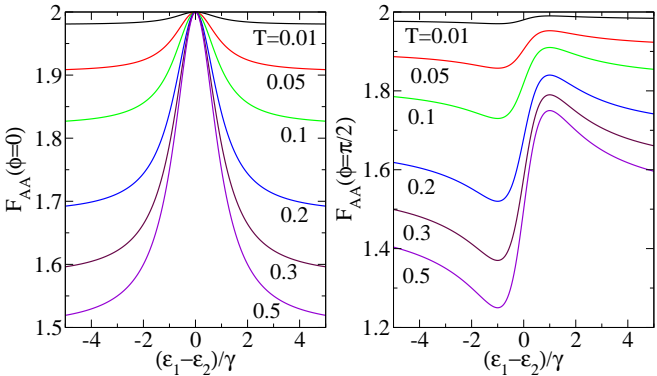}}
\caption{The Fano factor for the auto-correlations $F_{AA}=F_{BB}$ for
  phase difference $\phi=0$ (left panel) $\phi=\pi/2$ (right panel)
  and as a function of the normalized energy difference
  $(\varepsilon_1-\varepsilon_2)/\gamma$ for various beam-splitter
  transparencies.}
\label{fig3}
\end{figure}

The Fano factor as a function of energy difference
$\varepsilon_1-\varepsilon_2$ is plotted in Figs. \ref{fig3} and
\ref{fig4} for several values of the transparency of the beam
splitter.
For zero phase difference $\phi=0$, the Fano factor for the
cross-correlations shows a dip for aligned resonant levels. At
$\varepsilon_1-\varepsilon_2=0$, the Fano factor is zero, independent
on the beamsplitter transparency $T$. This is a sibnature of perfect
bunching. For finite phase-difference $\phi\neq 0$, the Fano factor
becomes asymmetric in $\varepsilon_1-\varepsilon_2$, showing a
Fano-shaped resonance, with the minimum shifted away from
$\varepsilon_1=\varepsilon_2$.

The Fano factor for the auto-correlations, for $\phi=0$, show a
corresponding peak for aligned resonant levels, reaching $2$ for
$\varepsilon_1=\varepsilon_2$. For finite phase-difference $\phi\neq
0$, the Fano factor becomes asymmetric, with the maximum Fano-factor
shifted away from $\varepsilon_1=\varepsilon_2$.

\begin{figure}[h]
\centerline{\psfig{figure=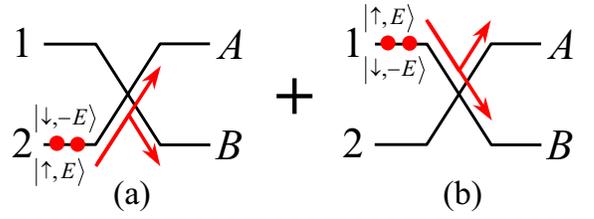,width=0.9\linewidth}}
\caption{Elementary scattering processes (shown at the beamsplitter)
contributing to the cross correlators $S_{AB}^{II}$. The two processes
(a) and (b) transport a pair of electrons $|\uparrow,E\rangle$ and
$|\downarrow,-E\rangle$ from the superconductor to the reservoirs $A$
and $B$ respectively.}
\label{figAB2}
\end{figure}

We point out that similarly to case I, the integrand of the cross
correlators can be understood by considering the basic two-particle
scattering processes. They are shown in Fig. \ref{figAB2}, the general
explanation is along the same line as for process I, discussed above.

\section{Discussion and Conclusions.}
In conclusion, we have investigated the spin entanglement in the
superconductor-quantum dot system proposed by Recher, Sukhorukov and
Loss [\onlinecite{Recher}]. Using a formal scattering theory we have
calculated the wavefunction of the electrons emitted by the entangler
and found that it is a superposition of spin-singlets at different
energies, a two particle wavepacket. Both the wavefunction for the two
electrons tunneling through different dots, creating the desired
nonlocal EPR-pair, as well as the wavefunction for the two electrons
tunneling through the same dot, were calculated.

The two electrons in the emitted pair collide in a beam-splitter
before exiting into normal reservoirs. Due to the symmetrical orbital
state, a consequence of the anti-symmetrical singlet spin-state, the
electrons tunneling through different dots show a tendency to
bunch. This bunching can be detected via the current correlations. It
was found that the amount of bunching depends on the position of the
single particle levels in the dots as well as on the scattering
properties of the beam splitter. Importantly, the magnitude of the
bunching dependent term in the cross correlations was found to be of
the same order as the bunching independent term, implying that an
experimental detection of the bunching, and thus indirectly the
spin-singlet entanglement, is feasible.

The current correlators for electrons tunneling through different dots
were found to be insensitive to orbital dephasing. Spin dephasing, on
the contrary, tends to randomize the spin state, leading to a mixed
spin-state with a finite fraction of triplets. Since singlet and
triplet spin states give rise to a bunching and anti-bunching behavior
respectively, when colliding in the beam-splitter, strong dephasing
will suppress the bunching behavior and eventually cause a crossover
to anti-bunching. To quantify the entanglement in the presence of spin
dephasing, we have derived an expression for the concurrence in terms
of the Fano factors. In addition, via the current correlations, it is
not possible to distinguish between entangled and non-entangled
spin-triplet states, since all triplets show the same bunching
behavior. This implies that the method of detecting spin entanglement
via current correlations in the beam-splitter geometry has a
fundamental limitation compared to the experimentally more involved
Bell Inequality test.

We have also investigated the current correlations in the case where
the two electrons tunnel through the same dot. The wavefunction was
found to be a linear superposition of states for the pair tunneling
through dots 1 and 2. The cross- and auto correlators are sensitive to
the position of the single-particle levels in the dots, however in a
different way than for tunneling through different dots. Moreover, the
correlators were found to be dependent on the scattering phases,
providing a way to distinguish between the two tunneling processes by
modulating the phase.

\section{Acknowledgements.}
We acknowledge discussions with Guido Burkard. The work was supported
by the Swiss National Science Foundation and the Swiss network for
Materials with Novel Electronic Properties.

\appendix
\section{}

In the presence of spin decoherence, the state of the pair of
electrons emitted through different dots can be described by a density
matrix $\rho$, which can be written as
\begin{eqnarray}
\rho &=&\left(\int dE |A(E)|^2\right)^{-1} \sum_{q,q'}\rho_{qq'}\int dE dE' \nonumber \\
&\times& A(E)A^*(E') |\Psi_q(E)\rangle\langle\Psi_{q'}(E')|
\label{densmat}
\end{eqnarray} 
noting that the normalization gives $\sum_q\rho_{qq}=1$. The index $q$
runs over the states in the singlet-triplet basis
$\{q\}=\{S,T_0,T_+,T_-\}$, i.e.
\begin{eqnarray}
|\Psi_S(E)\rangle&=&\frac{1}{\sqrt{2}}\left[b_{1\uparrow}^{\dagger}(E)b_{2\downarrow}^{\dagger}(-E)
-b_{1\downarrow}^{\dagger}(E)b_{2\uparrow}^{\dagger}(-E)\right]|0\rangle
\nonumber \\
|\Psi_{T_0}(E)\rangle&=&\frac{1}{\sqrt{2}}\left[b_{1\uparrow}^{\dagger}(E)b_{2\downarrow}^{\dagger}(-E)
+b_{1\downarrow}^{\dagger}(E)b_{2\uparrow}^{\dagger}(-E)\right]|0\rangle
\nonumber \\
|\Psi_{T_+}(E)\rangle&=&b_{1\uparrow}^{\dagger}(E)b_{2\uparrow}^{\dagger}(-E)|0\rangle
\nonumber \\
|\Psi_{T_-}(E)\rangle&=&b_{1\downarrow}^{\dagger}(E)b_{2\downarrow}^{\dagger}(-E)|0\rangle
\label{allstates}
\end{eqnarray}
The coefficients $\rho_{qq'}$ depend in general on the nature and the
strength of the spin decoherence. As pointed out in the text, only
energy independent spin decoherence is considered, and consequently
the coefficients $\rho_{qq'}$ are independent on energy.

The current operators conserve the individual spins. As a consequence,
the off-diagonal elements of $\rho$ do not contribute to the noise
correlators. As discussed in the text, all triplets contribute equally
to the correlators. Since the singlet and triplet states contribute
with opposite sign to the last term in Eqs. (\ref{crossfano}) and
(\ref{autofano}), the effect of spin-decoherence on the Fano factors
can be incorporated by renormalizing
$|H(\varepsilon_1-\varepsilon_2)|^2\rightarrow(2\rho_{SS}-1)|H(\varepsilon_1-\varepsilon_2)|^2$,
with the renormalization factor expressed in terms of $\rho_{SS}$ only
(using $\rho_{SS}+\rho_{T_0T_0}+\rho_{T_+T_+}+\rho_{T_-T_-}=1$), the
weight of the singlet component in $\rho$.

It is a difficult (and in general not analytically tractable) problem
to evaluate the entanglement of the full density matrix, since $\rho$
contains information about both the energy-dependent orbital part of
the state as well as the spin-part. In particular, due to the
continous energy variable, the dimension of $\rho$ is infinite. Here,
we are however only interested in the spin entanglement of $\rho$. To
determine the spin entanglement one has to consider measurement
schemes where the observables $O$ are sensitive only to the spin part of
$\rho$. Such observables satisfy the property
\begin{eqnarray}
&&\int
dEdE'A(E)A^*(E')\langle \Psi_{q}(E)|O| \Psi_{q'}(E')\rangle \nonumber \\
&&=
\langle \Psi_{q}|O_{\sigma}|\Psi_{q'}\rangle \int dE |A(E)|^2
\label{noendep}
\end{eqnarray}
where $|\Psi_{q'}\rangle$ are given from $|\Psi_{q'}(E)\rangle$ in
Eq. (\ref{allstates}) by removing the energy dependence, e.g.
$|\Psi_{T_+}\rangle=b_{1\uparrow}^{\dagger}b_{2\uparrow}^{\dagger}|0\rangle$.
The operator $O_{\sigma}$ is a function of the energy independent
$b$-operators. Using the property in Eq. (\ref{noendep}) we can write
\begin{eqnarray}
\langle O \rangle&=&\mbox{tr}[\rho O]=\left(\int dE |A(E)|^2\right)^{-1}
\sum_{q,q'}\rho_{qq'} \nonumber \\
&\times&\int dEdE'A(E)A^*(E')\langle \Psi_{q}(E)|O| \Psi_{q'}(E')\rangle
\nonumber \\
&=&\sum_{q,q'}\rho_{qq'}\langle
\Psi_{q}|O_{\sigma}|\Psi_{q'}\rangle \equiv \mbox{tr}[\rho_{\sigma} O_{\sigma}].
\end{eqnarray}
The $4\times4$ spin density matrix $\rho_{\sigma}$ is thus
\begin{eqnarray}
\rho_{\sigma}=\sum_{q,q'}\rho_{qq'}|\Psi_{q}\rangle\langle
\Psi_{q'}|. 
\end{eqnarray}
It is straightforward to show that for the special $\rho$ for aligned
dot levels $\varepsilon_1=\varepsilon_2$, the current correlators in
Eq. (\ref{noise}) are insensitive to the wave-packet structure of
$\rho$. In this case, $\rho_{\sigma}$ is directly obtained from $\rho$
by tracing over energies. More generally, independent of
$\varepsilon_1,\varepsilon_2$, the spin current correlators between
lead $1$ and $2$ (i.e. in the absence of the beamsplitter) are
insensitive to the wave-packet structure of $\rho$. These latter
correlators can be used to test a Bell Inequality, along the lines of
Ref. [\onlinecite{BI,Sam1}].

Our interest is thus to investigate the entanglement of
$\rho_{\sigma}$, conveniently expressed in terms of the
concurrence.\cite{Wootters} The concurrence $C$ is defined as
\begin{eqnarray}
C(\rho_{\sigma})=\mbox{max}\left\{0,\sqrt{\lambda_1}-\sqrt{\lambda_2}-\sqrt{\lambda_3}-\sqrt{\lambda_4}\right\}
\end{eqnarray}   
where the $\lambda_i$s are the real and positive eigenvalues, in
decreasing order, of $\rho_{\sigma}\tilde \rho_{\sigma}$. The matrix
$\tilde \rho_{\sigma}$ is defined as
\begin{eqnarray}
\tilde \rho_{\sigma}=(\sigma_y \otimes \sigma_y)
\rho_{\sigma}^*(\sigma_y \otimes \sigma_y)
\label{rhotilde}
\end{eqnarray}   
where $\sigma_y$ are Pauli matrices, rotating locally the spins in
lead $1$ and $2$ respectively. Importantly, in Eq. (\ref{rhotilde}),
the density matrix $\rho_{\sigma}$ is written in the spin-up/spin-down
basis,
i.e. $b_{1\uparrow}^{\dagger}b_{2\downarrow}^{\dagger}|0\rangle$ etc.
The concurrence is $C=0$ for an unentangled state and $C=1$ for a
state which is maximally entangled.

To determine $C(\rho_{\sigma})$, full information of $\rho_{\sigma}$
is needed. In the approach taken here, investigating the spin
entanglement via a beam-splitter and current correlators, one can
however not determine all elements of the density matrix
$\rho_{\sigma}$. As a consequence, the spin entanglement of the
emitted pair can not be determined precisely. It is nevertheless
possible, following the ideas of Burkard and Loss, \cite{Burk03} to
obtain a lower bound for the spin entanglement.

To obtain the lower bound, we first note two important properties of
$C(\rho_{\sigma})$: (i) $C(\rho_{\sigma})$ is invariant under local
rotations,\cite{Wootters} i.e. $C(\bar \rho_{\sigma})=C(\rho_{\sigma})$ for $\bar
\rho_{\sigma}=(U_1\otimes U_2)\rho_{\sigma}(U_2^{\dagger}\otimes
U_1^{\dagger})$, where $U_1$ and $U_2$ are unitary $2\times2$ matrices
acting locally on the spins in lead $1$ and $2$ respectively. (ii)
$C(\rho_{\sigma})$ is a convex function,\cite{Uhlmann} $\sum_{i}p_iC(\rho_i)\geq
C(\sum_ip_i\rho_i)$, i.e. for a density matrix $\rho_{\sigma}=\sum_i
p_i \rho_i$, with $\sum_i p_i=1$, the entanglement of the total
density density matrix is smaller than or equal to the weighted
entanglement of the parts (a consequence of information being lost
when adding density matrices).

Consider then the density matrix $\rho_W$ obtained by averaging
$\rho_{\sigma}$ with respect to all possible local rotations $U
\otimes U$, i.e. the same rotation in lead $1$ and $2$. Formally,
$\rho_W=(\langle U \otimes U) \rho_{\sigma} (U^{\dagger} \otimes
U^{\dagger})\rangle_U$ is calculated, where $\langle .. \rangle_U$
denotes an average with respect to $U$, uniformly distributed in the
group of unitary $2\times 2$ matrices. This gives the Werner state
\cite{Werner}
\begin{eqnarray}
&&\rho_W=\rho_{SS}|\Psi_{S}\rangle\langle \Psi_{S}|+\frac{1-\rho_{SS}}{3}\nonumber \\
&\times&\left(|\Psi_{T_0}\rangle\langle \Psi_{T_0}|+|\Psi_{T_+}\rangle\langle \Psi_{T_+}|+|\Psi_{T_-}\rangle\langle\Psi_{T_-}|\right)
\label{wernerstate}
\end{eqnarray}   
where we note that the singlet component is unaffected by the rotation
$U\otimes U$. Importantly, the entanglement of the Werner state is a
function of the singlet coefficient $\rho_{SS}$ only.  Using the two
properties (i) and (ii) of the entanglement stated above, we can write
\begin{eqnarray}
C(\rho_W)&=&C[\langle (U \otimes U) \rho_{\sigma} (U^{\dagger} \otimes
U^{\dagger}) \rangle_U ] \nonumber \\ &\leq& \langle C[(U \otimes U)
\rho_{\sigma} (U^{\dagger} \otimes U^{\dagger})] \rangle_U \nonumber \\
&=&\langle C(\rho_{\sigma}) \rangle_U= C(\rho_{\sigma})
\end{eqnarray}   
This shows that the concurrence of the Werner state $C_W=C(\rho_W)$
provides a lower bound for the entanglement of the full spin state
$C(\rho_{\sigma})$. The concurrence of the Werner state is
$C_W=\mbox{max} \{2\rho_{SS}-1,0\}$. The renormalization of the Fano
factors in Eqs. (\ref{crossfano}) and (\ref{autofano}) due to spin
decoherence can now simply be written
$|H(\varepsilon_1-\varepsilon_2)|^2\rightarrow
C_{W}|H(\varepsilon_1-\varepsilon_2)|^2$ where $C_{W}\geq 0$ is a
lower bound for the concurrence of the spin state in the presence of
decoherence. This is Eq. (\ref{renorm}) in the text.

\end{document}